\documentclass[prd,twocolumn,preprintnumbers,superscriptaddress,nofootinbib]{revtex4}
\usepackage[a4paper, hdivide={1.91cm,,1.165cm}, vdivide={1.83cm,,3.6cm}]{geometry}

\usepackage{amstext,amssymb}
\usepackage{amsmath}
\usepackage{graphicx}
\usepackage[hyperfootnotes=true]{hyperref}
\usepackage{xspace}
\usepackage{color}
\usepackage{units}
\usepackage{slashed} 

\begin{document}

\title{Universal Seesaw and $0\nu\beta\beta$ in new 3331 Left-Right Symmetric Model}
\author{Debasish Borah}
\email{dborah@iitg.ernet.in}
\affiliation{Department of Physics, Indian Institute of Technology Guwahati, Assam 781039, India}

\author{Sudhanwa Patra}
\email{sudha.astro@gmail.com}
\affiliation{Center of Excellence in Theoretical and Mathematical Sciences, \\
 \hspace*{-0.0cm} Siksha 'O' Anusandhan University, Bhubaneswar-751030, India}
\begin{abstract}
We consider a class of left-right symmetric model with enlarged gauge group $SU(3)_c \times SU(3)_L \times SU(3)_R \times U(1)_X$ 
without having scalar bitriplet. In the absence of scalar bitriplet, there is no Dirac mass term for fermions including usual quarks and leptons. 
We introduce new isosinglet vector-like fermions so that all the fermions get their masses through a universal seesaw mechanism. We extend our 
discussion to neutrino mass and its implications in neutrinoless double beta decay ($0\nu\beta\beta$). We show that for TeV scale $SU(3)_R$ gauge bosons, the heavy-light neutrino mixing contributes dominantly to $0\nu\beta\beta$ that can be observed at ongoing experiments. Towards the end we also comment on different 
possible symmetry breaking patterns of this enlarged gauge symmetry to that of the standard model. 

\end{abstract}
\maketitle 
\section{Introduction} 
\label{sec:intro}
The Standard Model (SM) of particle physics has been the most successful phenomenological theory specially after the discovery of its last missing piece, the Higgs boson at the Large Hadron Collider (LHC) back in 2012 with subsequent null results for Beyond Standard Model (BSM) searches. However, the SM fails to address several observed phenomena as well as theoretical questions. For example, it fails to explain the sub-eV neutrino mass \cite{PDG, T2K, chooz, daya, reno, minos}, the origin of parity violation in weak interactions and the origin of three fermion families. The first two questions can be naturally addressed within the framework of the Left-right symmetric model (LRSM) \cite{lrsm, lrsmpot}, one of the most widely studied BSM frameworks. These models not only explain tiny neutrino masses naturally through seesaw mechanism but also give rise to an effective parity violating SM at low energy through spontaneous breaking of a parity preserving symmetry at high scale. The conventional LRSM based on 
the gauge group $SU(3)_c \times SU(2)_L \times SU(2)_R \times U(1)_{B-L}$ can be enhanced to a more general LRSM based on the gauge group $SU(3)_c \times SU(3)_L \times SU(3)_R \times U(1)_{X}$ (or in short $3331$). The advantage of such an up gradation of the gauge symmetry is the ability of the latter in providing an explanation to the origin of three fermion families of the SM in addition to having other generic features of the LRSM. In such a model, the number of three fermion generations is no longer a choice, but a necessity in order to cancel chiral anomalies. In such models, where the usual lepton and quark representations are enlarged from a fundamental of $SU(2)_L$ in the SM to a fundamental of $SU(3)_L$, the number of generations must be equal to the number of colors in order to cancel the anomalies \cite{331ref}. This is in contrast with the SM or the usual LRSM where the gauge anomalies are canceled within each fermion generation separately. One can also build such a sequential $3331$ model by 
including additional chiral fermions. But since such a model does not explain the origin of three families from the anomaly cancellation point of view and contain non-minimal chiral fermion content, we stick to discussing a special type of non-sequential $3331$ model here.

There have been a few works \cite{Reig:2016tuk, Franco:2016hip, Reig:2016vtf, Dias:2010vt, 3331750} recently done 
within the framework of such $3331$ models with different motivations. Particularly from the origin of neutrino mass 
point of view, the work \cite{Reig:2016tuk} considered a scalar sector comprising of bitriplets plus sextets which 
gives rise to tiny neutrino masses through canonical type I \cite{ti} and type II \cite{tii0, tii} seesaw. Another 
recent work \cite{Reig:2016vtf} studied a specific $3331$ model with bitriplet and triplet scalar fields that can 
explain tiny neutrino masses through inverse \cite{inverse, linear} and linear seesaw mechanism \cite{linear}. 
The earlier work \cite{Dias:2010vt} considered effective higher dimensional operators to explain fermion masses 
in $3331$ models while the recent work \cite{3331750} studied the model and several of its variants from LHC 
phenomenology point of view. Here, we simply consider another possible way of generating fermion masses in $3331$ 
models through the universal seesaw mechanism \cite{VLQlr, univSeesawLR, univSeesawLR1} where all fermions acquire 
their masses through a common seesaw mechanism~\footnote{See Refs~\cite{Patra:2012ur,Dev:2015vjd,Deppisch:2016scs,Deppisch:2017vne} 
for implementation of universal seesaw mechanism 
for fermion mass generation within left-right symmetric model.}. Incorporating additional vector like fermion pairs corresponding 
to each fermion generation, we show that the correct fermion mass spectrum can be generated in such a model with a scalar 
sector where all of them transform as fundamentals under $SU(3)_{L,R}$ without the need of bi-fundamental and sextet 
scalars shown in \cite{Reig:2016tuk, Reig:2016vtf} for the implementation of different seesaw mechanism for neutrino 
masses. We also discuss the possibilities of light neutral fermions apart from sub-eV active neutrinos, their role 
in neutrinoless double beta decay $(0\nu \beta \beta)$ and different possible symmetry breaking chains of the gauge 
symmetry $SU(3)_c \times SU(3)_L \times SU(3)_R \times U(1)_{X}$ to that of the SM. We show that for TeV scale $SU(3)_R$ gauge bosons, the right handed neutrinos are constrained to lie around the keV mass regime having interesting consequences for $0\nu \beta \beta$. We find that although the pure heavy neutrino contribution to $0\nu \beta \beta$ remains suppressed compared to the one from light neutrinos, the heavy-light neutrino mixing which can be quite large in this model without any fine-tuning, gives a large contribution to $0\nu \beta \beta$ keeping it within experimental reach.

This letter is organized as follows. In section \ref{sec:model} we briefly discuss the model 
with the details of the particle spectrum, fermion masses via universal seesaw and gauge boson 
masses. In sections \ref{sec:0nubb} and \ref{sec:0nubb2} we discuss the contributions to $0\nu \beta \beta$ from purely light (heavy) neutrinos and heavy-light neutrino mixing respectively. Finally we discuss about different possible 
symmetry breaking chains in section \ref{sec:symbreak} and then conclude in section \ref{sec:conclude}.

\section{The model framework} 
\label{sec:model}
\subsection{Particle Spectrum}
The usual fermions transform under $SU(3)_c \times SU(3)_L \times SU(3)_R \times U(1)_X$ as
\begin{eqnarray}
 \Psi_{aL} &= \begin{pmatrix}\nu_{aL} \\ \ell^-_{aL} \\ \xi^q_{aL} \end{pmatrix}\, , & 
\Psi_{aR} = \begin{pmatrix}\nu_{aR} \\ \ell^-_{aR} \\ \xi^q_{aR} \end{pmatrix}\, ,\nonumber \\[1mm]
 Q_{m L} &= \begin{pmatrix} d_{\alpha L} \\ u_{\alpha L} \\ J^{-q-1/3}_{\alpha L} \end{pmatrix}\, , & 
Q_{m R} = \begin{pmatrix} d_{\alpha R} \\ u_{\alpha R} \\ J^{-q-1/3}_{\alpha R} \end{pmatrix}\,, \nonumber \\[1mm]
 Q_{3 L} &= \begin{pmatrix} u_{3 L} \\ d_{3 L} \\ J^{q+2/3}_{3 L} \end{pmatrix}\, , & 
Q_{3 R} = \begin{pmatrix} u_{3 R} \\ u_{3 R} \\ J^{q+2/3}_{3 R} \end{pmatrix} \,.
\end{eqnarray}
with a=1,2,3 whereas m =1,2. 

The transformation of the fields under the gauge symmetry $SU(3)_c \times SU(3)_L \times SU(3)_R \times U(1)_X$  are given in table \ref{table1}

\begin{center}
\begin{table}[b!]
\caption{Particle Content of the Model}
\begin{tabular}{|c|c|}
\hline
Particle & $SU(3)_c \times SU(2)_L \times SU(3)_R \times U(1)_X$  \\
\hline
$ Q_{mL} $ & $(3,3^*,1,-\frac{q}{3})$  \\
$ Q_{mR} $ & $(3,1,3^*,-\frac{q}{3})$\\
$ Q_{3L} $ & $(3,3,1,\frac{q+1}{3})$  \\
$ Q_{3R} $ & $(3,1,3,\frac{q+1}{3})$\\
$ \Psi_{aL} $ & $(1,3,1,\frac{q-1}{3})$  \\
$ \Psi_{aR} $ & $(1,1,3,\frac{q-1}{3})$\\
$ U_{L,R} $ & $(3,1,1,\frac{2}{3})$  \\
$ D_{L,R} $ & $(3,1,1,\frac{2}{3})$\\
$ E_{L,R} $ & $(1,1,1,-1)$  \\
$ N_{L,R} $ & $(1,1,1,0)$\\
\hline
$ \chi_L$ & $(1,3, 1,-\frac{q+2}{3})$\\
$ \chi_R$ & $(1,1, 3,-\frac{q+2}{3})$\\
$ \phi_L$ & $(1,3, 1,\frac{1-q}{3})$\\
$ \phi_R$ & $(1,1, 3,\frac{1-q}{3})$\\
\hline
\end{tabular}
\label{table1}
\end{table}
\end{center}

Assuming $q=0$, if the neutral components of the scalar fields acquire their vacuum expectation value (vev) as 
\begin{equation}
\langle \chi_{L} \rangle \equiv \frac{1}{\sqrt{2}} \begin{pmatrix}   0 \\  v_{1L}  \\ 0  \end{pmatrix}, \;\; \langle \phi_{L} \rangle \equiv \frac{1}{\sqrt{2}} \begin{pmatrix}   v_{2L} \\  0  \\ \omega_{L}  \end{pmatrix} \nonumber 
\end{equation} 
\begin{equation}
\langle \chi_{R} \rangle \equiv \frac{1}{\sqrt{2}} \begin{pmatrix}   0 \\  v_{1R}  \\ 0  \end{pmatrix}, \;\; \langle \phi_{R} \rangle \equiv \frac{1}{\sqrt{2}} \begin{pmatrix}   v_{2R} \\  0  \\ \omega_{R}  \end{pmatrix}
\label{scalarvev}
\end{equation} 

\subsection{Fermion Mass}
The Yukawa Lagrangian can be written as 
\begin{align}
\hspace*{-0.5cm}\mathcal{L}_Y &= - \left[Y_D\right]_{ma} \left( \overline{Q}_{mL} \phi_L D_{aR} + \overline{Q}_{mR} \phi_R D_{aL}  \right) - \left[M_D\right]_{ab}  \overline{D}_{aL} D_{bR} \nonumber \\
 &\hspace*{-0.5cm}\quad - \left[Y_U\right]_{ma} \left( \overline{Q}_{mL} \chi_L U_{aR} + \overline{Q}_{mR} \chi_R U_{aL}  \right) - \left[M_U\right]_{ab}  \overline{U}_{aL} U_{bR} \nonumber \\
 &\hspace*{-0.5cm}\quad - \left[Y^\prime_D\right]_{3a} \left( \overline{Q}_{3L} \chi^{*}_L D_{aR} + \overline{Q}_{3R} \chi^{*}_R D_{aL}  \right) \nonumber \\
 &\hspace*{-0.5cm}\quad - \left[Y^\prime_U\right]_{3a} \left( \overline{Q}_{3L} \phi^{*}_L U_{aR} + \overline{Q}_{3R} \phi^{*}_R U_{aL}  \right)  \nonumber \\
 &\hspace*{-0.5cm}\quad - \left[Y_E\right]_{ab} \left( \overline{\Psi}_{aL} \chi^{*}_L E_{bR} + \overline{\Psi}_{mR} \chi^{*}_R E_{bL}  \right) - \left[M_E\right]_{ab}  \overline{E}_{aL} E_{bR} \nonumber \\
 &\hspace*{-0.5cm}\quad - \left[Y_N\right]_{ab} \left( \overline{\Psi}_{aL} \phi^{*}_L N_{bR} + \overline{\Psi}_{mR} \phi^{*}_R N_{bL}  \right) - \left[M_{LR}\right]_{ab}  \overline{N}_{aL} N_{bR} \nonumber \\
 &\hspace*{-0.5cm}\quad - \left[M_{LL}\right]_{ab} N_{aL} N_{bL} - \left[M_{RR}\right]_{ab} N_{aR} N_{bR} +\text{h.c.} 
\end{align}
%
\begin{figure}[h!]
\includegraphics[width=1.05\linewidth]{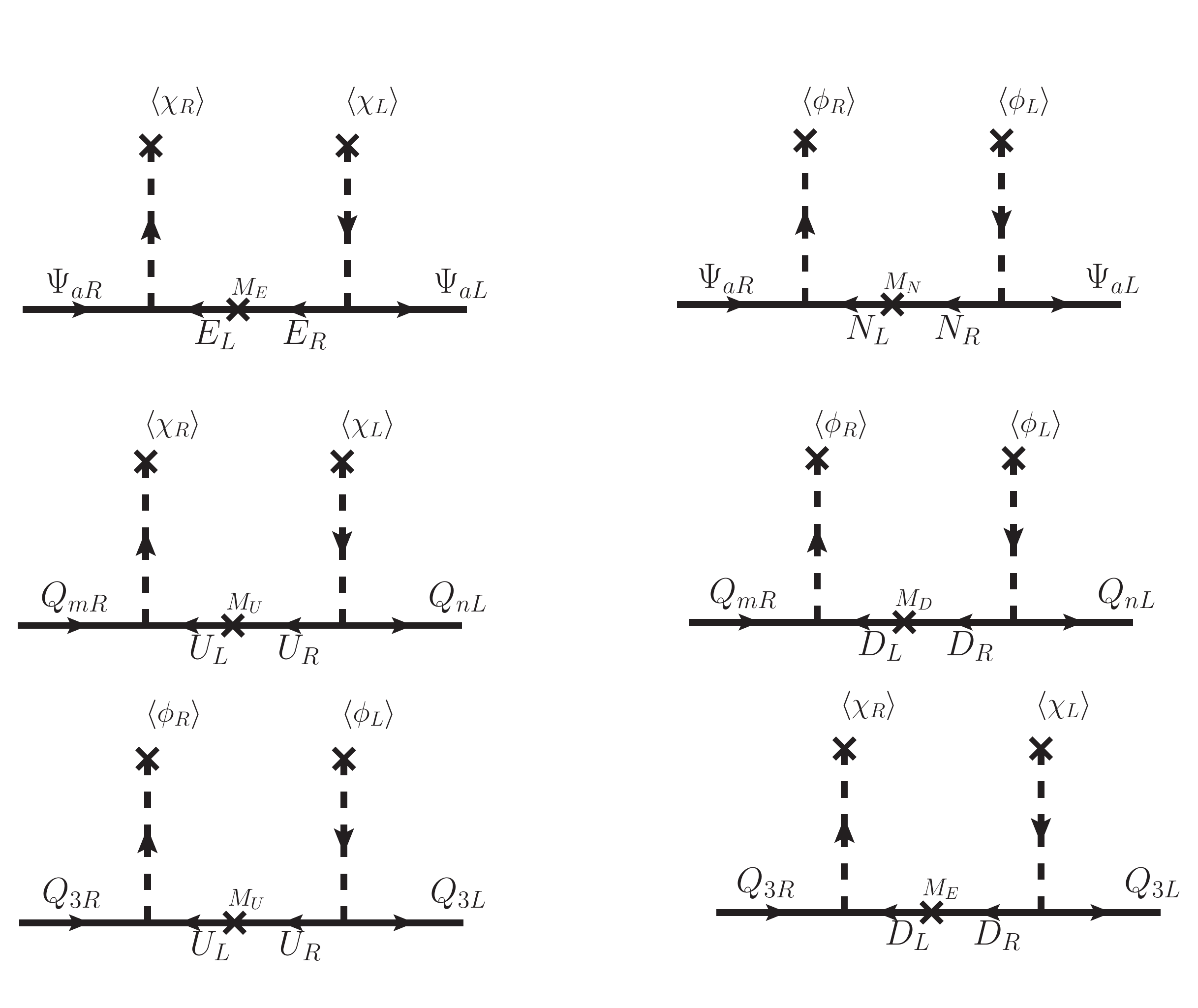}
\caption{Feynman diagram for Dirac mass of fermions within 
         $SU(3)_c \times SU(3)_L \times SU(3)_R \times U(1)_X$ model.}
\label{feyn1}
\end{figure}
After integrating out the heavy fermions, we can write down the effective Yukawa terms for 
charged fermions of the standard model as follows
\begin{align}
&y_u = Y_U \frac{v_{1R}}{M_U} Y^T_U, \nonumber \\
&y_d = Y^{}_D  \frac{v_{2R}}{M_D} Y^{T}_D, \nonumber \\
&y_e = Y_E \frac{v_{1R}}{M_E} Y^{T}_E
\end{align}

Similarly, the heavy neutral singlet fields $N_{L,R}$ can be integrated out to generate 
the effective mass matrix of neutrinos $\nu_L, \nu_R$ which contains a Dirac mass term and two Majorana 
mass terms. The effective Dirac mass as well as Majorana mass terms are given by  
\begin{eqnarray}
&& M_D = Y^{}_{N} \frac{1}{M_{RR}} M^T_{LR} \frac{1}{M_{RR}} Y^{T}_{N} v_{2L} v_{2R}\, , \nonumber \\
&& M_L = Y^{}_{N} \frac{1}{M_{RR}} Y^{T}_{N} v^2_{2L}\, , \nonumber \\
&& M_R = Y^{}_{N} \frac{1}{M_{RR}} Y^{ T}_{N} v^2_{2R}\, .
\label{numass5}
\end{eqnarray}

There are additional neutrino leptons $\xi_L,\xi_R$ which acquire Dirac and Majorana masses similar to $\nu_{L,R}$ shown above. They are given by 
\begin{eqnarray}
&& M_{\xi_D} = Y^{}_{N} \frac{1}{M_{RR}} M^T_{LR} \frac{1}{M_{RR}} Y^{T}_{N} \omega_{L} \omega_{R}\, , \nonumber \\
&& M_{\xi_L} = Y^{}_{N} \frac{1}{M_{RR}} Y^{T}_{N} \omega^2_{L}\, , \nonumber \\
&& M_{\xi_R} = Y^{}_{N} \frac{1}{M_{RR}} Y^{ T}_{N} \omega^2_{R}\, .
\label{eq:numass_rel}
\end{eqnarray}
\begin{figure}[h!]
\includegraphics[width=1.05\linewidth]{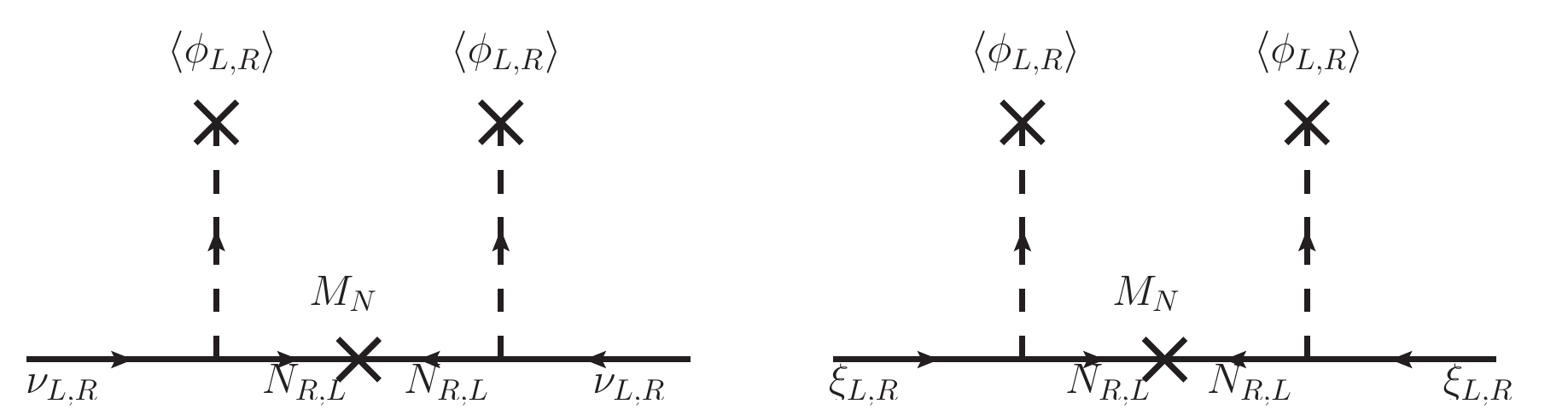}
\caption{Feynman diagram for Majorana mass of neutral fermions within 
         $SU(3)_c \times SU(3)_L \times SU(3)_R \times U(1)_X$ model.}
\label{feyn2}
\end{figure}
\begin{figure}[h!]
\includegraphics[width=0.99\linewidth]{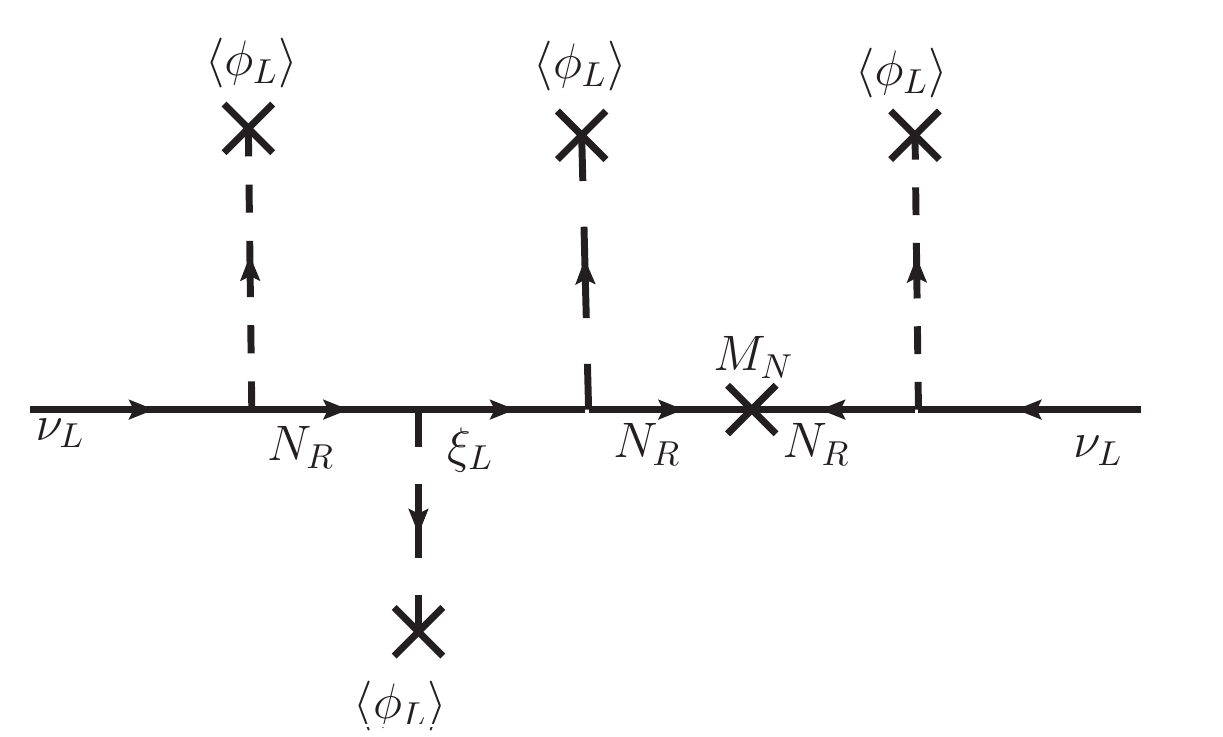}
\caption{Feynman diagram for new contribution to the Majorana mass of light neutrinos 
        within $SU(3)_c \times SU(3)_L \times SU(3)_R \times U(1)_X$ model.}
\label{feyn3}
\end{figure}
The origin of the Dirac masses can be understood from the mass diagrams shown in figure \ref{feyn1} whereas the Majorana mass diagrams are given in figure \ref{feyn2}. Apart from these, the light neutrinos also receive non-leading contribution to their masses from the diagram shown in figure \ref{feyn3}. The contribution of this diagram can be written as
\begin{equation}
M^{\prime}_L = Y^{}_{N} \frac{1}{M_{RR}} Y^{T}_{N} \frac{1}{M_{\xi_L}} Y^{}_{N} \frac{1}{M_{RR}} Y^{T}_{N} v^2_{2L} \omega^2_L
\end{equation}
Here we are assuming equality of left and right sector Yukawa couplings as well as masses 
$M_{LL}=M_{RR}=M_N$. The approximate scale of light neutrino mass matrix $M_L$ has to be less than 
$0.1$ eV which puts limit on model parameters $Y_N, M_{RR}$. For example, if $Y_N \simeq 0.01$ then $M_{RR} \geq 10^{10}~$GeV to keep $M_L \leq 0.1$ eV. Further lowering the scale of $M_{RR}$ will involve more fine-tuning in the Yukawa coupling $Y_N$.  In the limit of tiny $M_D$, the light neutrino mass solely originates from $M_L$ given in equation \eqref{numass5} and the mixing between heavy and light neutrinos can be neglected. In such a case, the light neutral lepton mass matrix can be written 
in the basis $(\nu_L, \nu_R)$ as
\begin{align}
	M_\nu= 
	\left(\begin{array}{cc}
		Y^{}_{N} \frac{1}{M_{RR}} Y^{T}_{N} v^2_{2L}  & 0  \\
   	    0 & Y^{}_{N} \frac{1}{M_{RR}} Y^{ T}_{N} v^2_{2R}
\end{array} \right) \,.
\label{eqn:numatrix}       
\end{align}
As a result of this particular structure of the mass matrix, both the mass eigenvalues for left-handed 
and right-handed neutrinos are proportional to each other. The two mixing matrices are related as
\begin{align}
  V_{\nu_R} = V_{\nu_L} \equiv U \,,
\label{eq:equality-VLR}
\end{align}
where $U$ is the Pontecorvo-Maki-Nakagawa-Sakata (PMNS) leptonic mixing matrix. 
For a representative sets of input model parameters like $v_L\approx 174~$GeV, $v_R \approx 10~$TeV, 
and $m_i\approx 0.1~$eV, the right-handed neutrino masses lie in the range of keV scale. 
Such light keV scale right handed neutrinos can also have very 
interesting implications for cosmology \cite{dblr16}. We leave such a detailed study of this model from cosmology 
point of view to an upcoming work.

\subsection{Gauge Boson Mass}
The relevant kinetic terms leading to gauge boson masses are given by
\begin{eqnarray}
\mathcal{L}_{\rm G.B.} &&\supset \left|\left(\frac{g_L}{2} \textbf{W}^{L}_\mu+g_X (-2/3) B_{\mu}\right) \chi_{L}\right|^2 \nonumber\\
  &&+ \left|\left(\frac{g_R}{2} \textbf{W}^{R}_\mu+g_X (-2/3) B_{\mu}\right) \chi_{R}\right|^2 \nonumber\\
  &&+ \left|\left(\frac{g_L}{2} \textbf{W}^{L}_\mu+g_X (1/3) B_{\mu}\right) \phi_{L}\right|^2 \nonumber\\
  &&+ \left|\left(\frac{g_R}{2} \textbf{W}^{R}_\mu+g_X (1/3) B_{\mu}\right) \phi_{R}\right|^2
 \label{eq:boson1}
\end{eqnarray}
where the factor $\textbf{W}^{L,R}_\mu$ is defined as
\begin{eqnarray}
\textbf{W}^{L,R}_\mu&&=\sum^8_{i=1}W^i_{L,R\hspace{0.3mm}\mu}\Lambda_i \nonumber \\
&&=
\left (\begin{array}{ccc}
W^3+\frac{1}{\sqrt{3}}W^8 & W^+ & V^{-q}\\
W^- & -W^3+\frac{1}{\sqrt{3}}W^8 & V^{\prime \hspace{0.1mm}-q-1}\\
V^{q} & V^{\prime \hspace{0.1mm}q+1} & -\frac{2}{\sqrt{3}}W^8\\
\end{array}\right )_{L,R} \nonumber
\end{eqnarray}
Using respective vev's for scalar fields shown in equation \eqref{scalarvev} we can derive the gauge boson masses for the present model. In the gauge boson spectrum, we have
\begin{itemize}
 \item One massless photon $A$,
 \item Four neutral gauge bosons $Z_{L,R}$, $Z^\prime_{L,R}$,
 \item Four charged gauge bosons $W^\pm_{L,R}$,
 \item Four gauge bosons with charge $q+1$, $X^{\pm(1+q)}_{L,R}$,
 \item Four gauge bosons with charge $q$, $Y^\pm_{L,R}$.
\end{itemize}

\section{$0\nu\beta\beta$ with purely light (heavy) neutrino contributions}
\label{sec:0nubb}
We find that the light sub-eV scale left-handed neutrinos ($\nu_L$) and the heavy right-handed neutrinos ($\nu_R$) 
with keV scale masses can give sizable contributions to neutrinoless double beta decay. Since the bitriplet scalar 
is absent in the present left-right symmetric $3331$ model, there are no Dirac mass term for light neutrinos at tree level. 
However one can eventually generate Majorana masses for $\nu_L$ and $\nu_R\equiv N_R$ through universal seesaw, 
see Fig.\ref{feyn2}. Such Majorana nature of neutrinos violate lepton number by two units and thus, contributes 
to $0\nu\beta\beta$ decay. The Feynman diagram for $0\nu\beta\beta$ decay is depicted in Fig.\ref{feyn4} due to exchange of left-handed as well as right-handed 
neutrinos. The corresponding Feynman amplitudes due to exchange of left-handed and right-handed neutrinos are given by 
\begin{eqnarray}
&&\mathcal{A}_{\nu_L} \propto G^2_F\, \frac{U^2_{e\,i}\, m_{i}}{p^2}  \nonumber \\
&&\mathcal{A}_{\nu_R} \propto G^2_F\, \left(\frac{M^2_{W_L}}{M^2_{W_R}}\right)\,
U^2_{e\,i}\, \frac{M_{i}}{p^2}
\end{eqnarray}

\begin{figure}[t!]
	\includegraphics[width=0.51\textwidth]{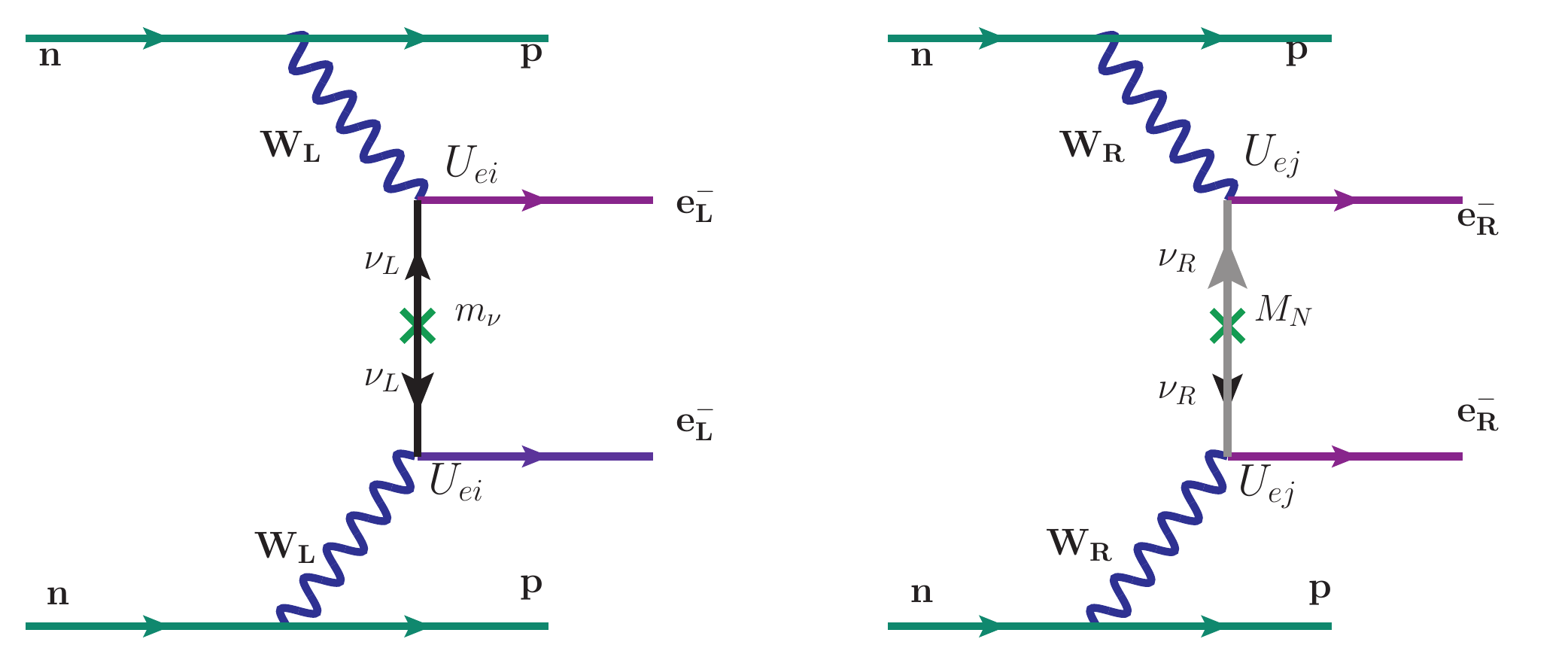}
	\caption{$0\nu\beta\beta$ decay diagrams due to exchange of light left-handed neutrinos with left-handed charged currents 
	and right-handed neutrinos with right-handed charged currents.} 
	\label{feyn4}
\end{figure}
The inverse half-life for a given isotope for $0\nu\beta\beta$ decay- due to exchange of left-handed 
light neutrinos via left-handed currents, and right-handed neutrinos via right-handed currents is given by
\begin{align}
 [T_{1/2}^{0\nu}]^{-1} \!=\! G_{01} 
 \left(|\mathcal{M}_\nu \eta_{\nu}|^2 + |\mathcal{M}_N \eta_N |^2 \right),
\end{align}
where $G_{01}$ is $0\nu\beta\beta$ phase space factor, $\mathcal{M}_i$ correspond to the 
nuclear matrix elements (NME) and $\eta_i$ is the corresponding dimensionless particle physics parameter. 
Since we have $M_i \approx 1-10~$MeV masses of right-handed neutrinos in the present model and satisfying 
$|M^2_i| \ll p^2$ where $p$ being the neutrino virtually momentum around $100~$MeV, the NMEs for right-handed neutrinos 
and left-handed neutrinos are same i.e, $\mathcal{M}_\nu = \mathcal{M}_N$.

\begin{table}[htb]
\begin{center}
\begin{tabular}{|c|c|}
\hline
$\eta_i$     &  {\bf Effective Mass Parameter} 
\rule{0pt}{2.1em}\\  [10pt]
\hline 
$\eta_{\nu_L} \approx \frac{1}{m_e}  \sum^{3}_{i=1} U^2_{ei}\, m_{i} $  &  
 $m^{\nu_L}_{\rm ee} \approx \sum^{3}_{i=1} U^2_{ei}\, m_{i} $ 
  \rule{0pt}{2.1em}\\  [10pt]
$\eta_{\nu_R} \approx 
	\frac{1}{m_e} \left(\frac{M_{W_L}}{M_{W_R}}\right)^4  
              \sum^{3}_{i=1} U^2_{ei} M_i$ &  
 $m^{\nu_R}_{\rm ee} \approx \left(\frac{M_{W_L}}{M_{W_R}}\right)^4  
              \sum^{3}_{i=1} U^2_{ei} M_i $            
  \rule{0pt}{2.1em}\\  [10pt]
\hline
\end{tabular}
\caption{Dimensionless particle physics parameters due to exchange of left-handed and right-handed 
         neutrinos and the corresponding effective mass parameters. }
\label{table:onubb-etamee}
\end{center}
\end{table}
\begin{figure*}[t!]
\includegraphics[width=0.48\textwidth]{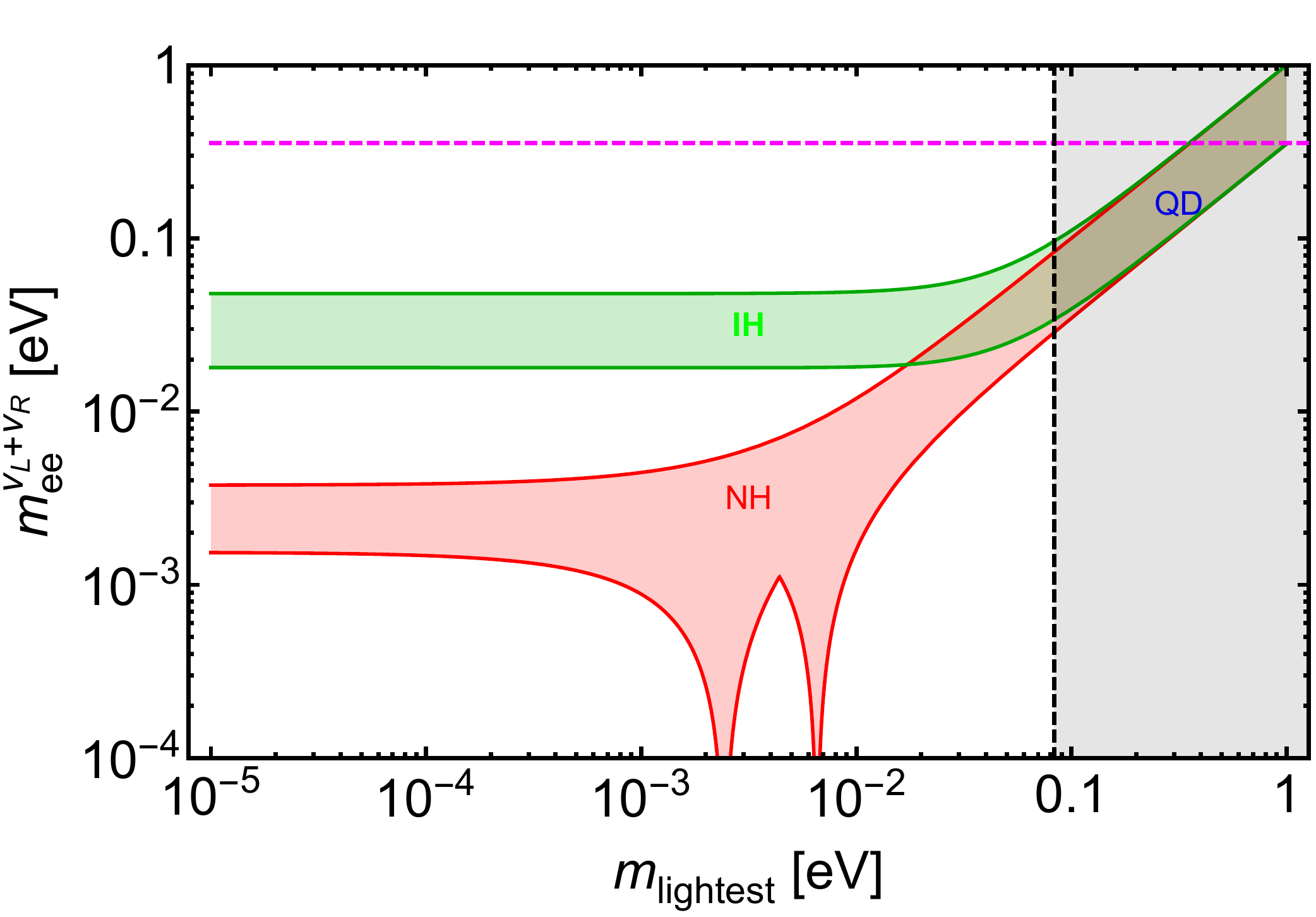}
\includegraphics[width=0.48\textwidth]{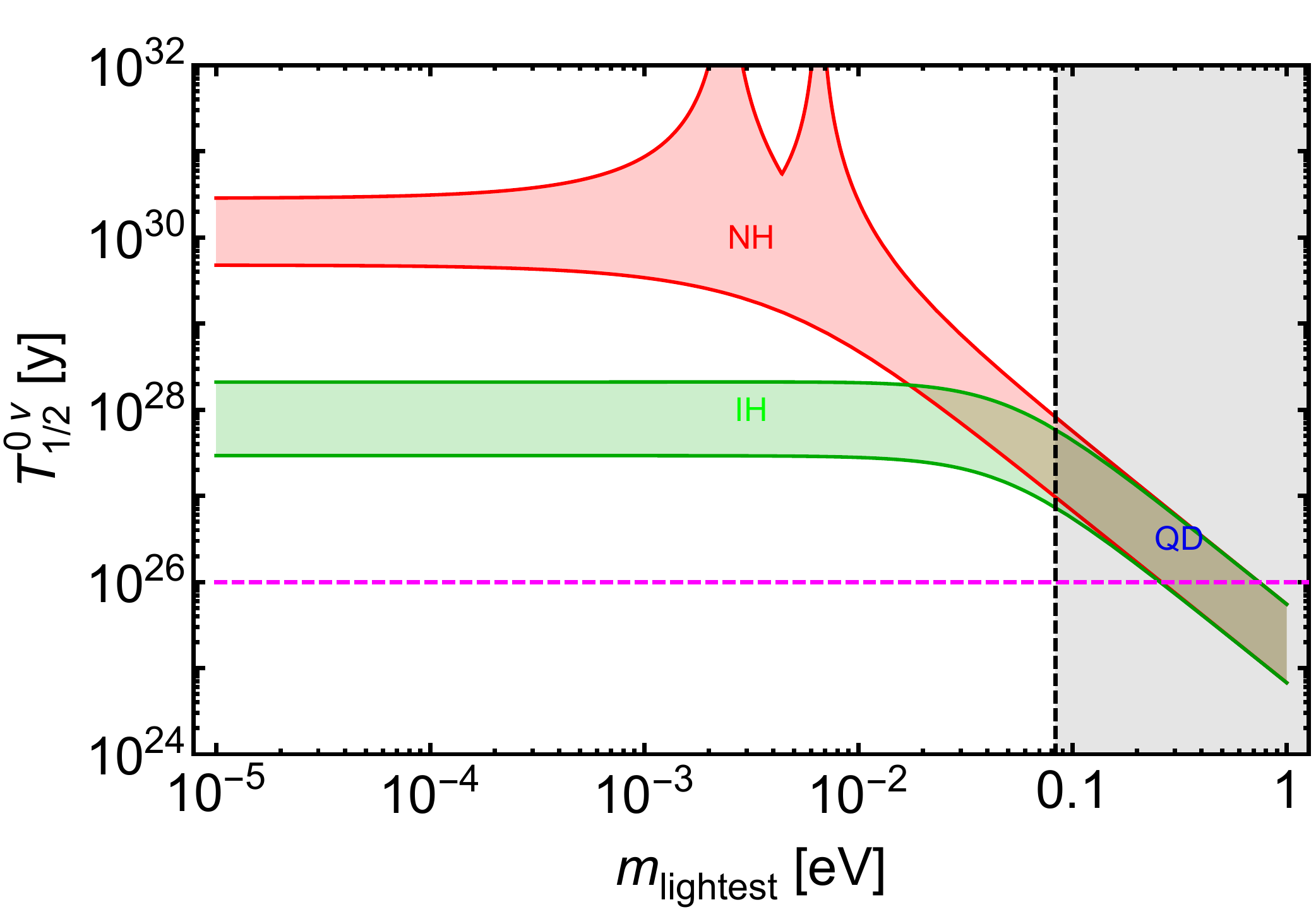}
\caption{Effective mass (left-panel) and Half-life (right-pannel) as a function of the lightest
         neutrino mass. We have used best-fit oscillation data~\cite{GonzalezGarcia:2012sz} and the Majorana phases are varied 
         between $[0,2\pi]$. The vertical shaded regions are excluded from cosmology~\cite{Ade:2015xua,Mertens:2015ila} 
         while the dashed horizontal line is for from KamLAND-Zen~\cite{KamLAND-Zen:2016pfg} bound. } 
\label{plot1}
\end{figure*}

\subsection{\hspace*{-0.2cm}Standard mechanism via left-handed neutrinos $\nu_L$}
The standard mechanism for neutrinoless double beta decay due to exchange of light 
left-handed neutrinos via left-handed currents gives dimensionless particle physics 
parameter as,
\begin{align}
\label{eta:nu} 
\mathcal{\eta}_{\nu_L} =\frac{1}{m_e}  \sum^{3}_{i=1} U^2_{ei}\, m_{i}
           = \frac{m^{\nu_L}_{\rm ee}}{m_e} \,. 
\end{align}
Here, $m_e$ is the electron mass. The effective mass parameter for standard mechanism is explicitly given by
\begin{align}
\label{eq:mee-std}
m^{\nu_L}_{\rm ee}
=\left| c^2_{12} c^2_{13} m_1 + s^2_{12} c^2_{13} m_2 e^{i\alpha} + s^2_{13} m_3 e^{i\beta} \right| \,,
\end{align}

\subsection{\hspace*{-0.2cm}New contribution from right-handed neutrinos $\nu_R$}
In the present left-right symmetric $3331$ model, we found that the right-handed neutrino mass lies 
around a few keV (for TeV scale $W_R$) which is much less than its momentum, $M_i \ll |p|$. Under this condition, 
the propagator simplifies in a similar way way as for the light neutrino exchange,
\begin{align}
   P_{R}\frac{\slashed{p}+M_i}{p^2-M_i^2}P_{R} \approx \frac{M_i}{p^2}\,.
\end{align}
This results dimensionless particle physics parameter $\eta_{\nu_R}$ due to exchange of right-handed 
neutrinos via right-handed currents as,
\begin{align}
	\eta_{\nu_R} \approx \frac{1}{m_e}\left(\frac{M_{W_L}}{M_{W_R}}\right)^4
             \sum_{i=1}^3 U^2_{ei} M_i \propto \eta_{\nu_L} \,.
\end{align} 
where the proportionality relation between $\eta_{\nu_R}$ and $\eta_{\nu_L}$ appears at the last 
step due to the proportionality between heavy and light neutrino mass matrix discussed above. 
After a little simplification, the effective mass parameter due to exchange of right-handed neutrinos 
can be expressed as,
\begin{align}
m^{\nu_R}_{\rm ee} \approx \left(\frac{M_{W_L}}{M_{W_R}}\right)^4 
             \sum_{i=1}^3 U^2_{ei} M_i \propto m^{\nu_L}_{\rm ee} \,.
\end{align} 
It is clear from eq.(\ref{eq:numass_rel}) that both light and heavy neutrino mass eigenvalues are 
proportional to each other as
\begin{align}
M_{i} \approx \frac{v^2_R}{v^2_L} m_i \,.
\end{align}
Thus, one can express $W_R$ mass as 
\begin{align}
M^2_{W_R} \approx \frac{1}{4}g^2_R v^2_R = \frac{1}{4} g^2_R v^2_L \frac{M_i}{m_i}  \,.
\end{align}
As a result, the effective Majorana mass parameter-- with $g_L\approx g_R$ and $M_W \approx \frac{1}{2} g_L v_L$-- is modified to
\begin{align}
m^{\nu_R}_{\rm ee} \approx \left(\frac{m_{1}}{M_{1}}\right)^2
             \sum_{i=1}^3 U^2_{ei} M_i \,.
\end{align}
Comparing this with the light neutrino contribution $m^{\nu_L}_{\rm ee} =  \sum^{3}_{i=1} U^2_{ei}\, m_{i}$, 
it is straightforward to estimate that the heavy neutrino contribution is suppressed by a factor of 
$m_i/M_i = (v_L/v_R)^2$ compared to the light neutrino contribution.
\subsection{Numerical Results}
The total contribution to inverse half-life for neutrinoless double beta decay for a given isotope 
in the present left-right symmetric $3331$ model due to exchange of left-handed as well as right-handed 
neutrinos is given by
\begin{align}
 [T_{1/2}^{0\nu}]^{-1} \!=\! G_{01} \left| \frac{\mathcal{M}_\nu}{m_e}\right|^2 
|m_{\rm eff}^{\rm ee}|^2,
\end{align}
where,
\begin{eqnarray}
|m^{\rm eff}_{\rm ee}|^2&=&|  U^2_{ei} \, m_i\ |^2 + \Bigg| \frac{M^4_{W_L}}{M_{W_R}^4} 
                          U_{\,ei}^2\, M_{i}   \Bigg|^2\,, \nonumber \\
                      &=& |m^{\nu_L}_{\rm ee}|^2 + |m^{\nu_R}_{\rm ee}|^2 \  
\label{eqn:mee-nuN}    
\end{eqnarray}

\subsubsection{For NH Pattern}
For normal hierarchical (NH) pattern of light neutrinos we consider the following 
mass structures for left-handed and right-handed neutrinos,
\begin{eqnarray}
&&m_1  = m_{\rm lightest}\,, \quad \quad  m_2 = \sqrt{m^2_1 + \Delta m^2_{\rm sol}}\, , \nonumber \\
&&m_3 = \sqrt{m^2_1 + \Delta m^2_{\rm sol} + + \Delta m^2_{\rm atm}}\,, \nonumber \\
&&M_{>} = M_3\, , \nonumber \\
&&M_{1} = \frac{m_{1}}{m_{3}} M_{3}, \quad M_{2} = \frac{m_{2}}{m_{3}} M_3\,. 
\end{eqnarray}
where $M_3$ is fixed around few keV range, as a result of choosing the $W_R$ mass scale at a few TeV. The analytic form for effective mass parameters due to 
exchange of right-handed neutrinos is given by
\begin{eqnarray}
& &m^{\nu_L}_{\rm ee} =\left| c^2_{12} c^2_{13} m_1 + s^2_{12} c^2_{13} m_2 e^{i\alpha} + s^2_{13} m_3 e^{i\beta} \right|\, , \nonumber \\
&&\hspace*{-0.8cm}|m^{\nu_R}_{\rm ee}|_{\rm NH} = \left(\frac{m_{1}}{M_{1}}\right)^2\,M_{3}\, \bigg| 
\frac{m_1}{m_3} c^2_{12} c^2_{13} +\frac{m_2}{m_3} s^2_{12} c^2_{13} e^{i\alpha} + s^2_{13}\,e^{i\beta} \bigg| \nonumber \\
\label{eqn:M} 
\end{eqnarray}

\subsubsection{For IH Pattern}
Similarly for inverse hierarchical (IH) pattern of light neutrinos, the masses for light left-handed and right-handed 
neutrinos are fixed as,
\begin{eqnarray}
&&m_3  = m_{\rm lightest}\,, \quad \quad  m_2 = \sqrt{m^2_1 + \Delta m^2_{\rm sol}}\, , \nonumber \\
&&m_3 = \sqrt{m^2_1 + \Delta m^2_{\rm sol} + + \Delta m^2_{\rm atm}}\,, \nonumber \\
&&M_{>} = M_2\, , \nonumber \\
&&M_{1} = \frac{m_{1}}{m_{2}} M_{2}, \quad M_{3} = \frac{m_{3}}{m_{2}} M_2\,. 
\end{eqnarray}
where we fixed the heaviest right-handed neutrino mass $M_2$ around few keV. The effective mass parameters due to 
exchange of right-handed neutrinos is given by
\begin{eqnarray}
& &m^{\nu_L}_{\rm ee} =\left| c^2_{12} c^2_{13} m_1 + s^2_{12} c^2_{13} m_2 e^{i\alpha} + s^2_{13} m_3 e^{i\beta} \right|\, , \nonumber \\
& &\hspace*{-0.7cm}|m^{\nu_R}_{\rm ee}|_{\rm IH}=\left(\frac{m_{1}}{M_{1}}\right)^2\,M_{2}\, \bigg| 
\frac{m_1}{m_2} c^2_{12} c^2_{13}+ s^2_{12} c^2_{13} e^{i \alpha}
           + \frac{m_3}{m_2} s^2_{13} \,e^{i \beta} \bigg|  \nonumber \\
\label{eqn:M} 
\end{eqnarray}

We have generated effective mass (left-panel) and half-life (right-pannel) with the variation of 
lightest neutrino mass, $m_{\nu_1}$ for NH and $m_{\nu_3}$ for IH as shown in Fig.\ref{plot1}. 
The green and red lines are for NH and IH pattern of light neutrinos applicable for both left- and 
right-panel. The horizontal dashed line is for bound from $0\nu\beta\beta$ experiments while the vertical 
dashed lines along with brown shaded regions are excluded from Planck limit. The present bound on half-life 
is $T_{1/2}^{0\nu}(^{136}\text{Xe}) > 1.07\times 10^{26}$~yr at 90\% C.L. from KamLAND-Zen~\cite{KamLAND-Zen:2016pfg}. 
It is found that the new physics contributions to $0\nu\beta\beta$ is very much suppressed and the standard 
mechanism due to exchange of light neutrinos are dominant. The total contribution to $0\nu\beta\beta$ is similar to the light neutrino contribution, saturating the experimental bound only near the quasi-degenerate regime, clearly visible from the plots shown in figure \ref{plot1}.
\section{$0\nu\beta\beta$ with heavy-light neutrino mixing}
\label{sec:0nubb2}
Though the purely heavy neutrino contribution to $0\nu \beta \beta$ remains suppressed, there can be sizable contributions from heavy-light neutrino mixing diagrams. This heavy-light neutrino mixing can also contribute to light neutrino masses through a type I seesaw formula which was ignored in the above discussion for simplicity. Such an assumption is valid for negligible heavy-light neutrino mixing. However, if we go beyond this simple assumption or equivalently consider $M_{LR} \approx M_{RR}$, then the Dirac as well as Majorana mass matrices for $\nu_L$ and $\nu_R \equiv N_R$ can be written as 
\begin{eqnarray}
&& M_D = Y^{}_{N} \frac{1}{M_{RR}} Y^{T}_{N} v_{2L} v_{2R}\, , \nonumber \\
&& M_L = Y^{}_{N} \frac{1}{M_{RR}} Y^{T}_{N} v^2_{2L}\, , \nonumber \\
&& M_R = Y^{}_{N} \frac{1}{M_{RR}} Y^{ T}_{N} v^2_{2R}\, .
\end{eqnarray}
Thus, the neutral lepton mass matrix in the basis $(\nu_L, N_R)$ as
\begin{eqnarray}
M_\nu &=& 
	\left(\begin{array}{cc}
		Y^{}_{N} \frac{1}{M_{RR}} Y^{T}_{N} v^2_{2L}  & Y^{}_{N} \frac{1}{M_{RR}} Y^{T}_{N} v_{2L} v_{2R} \\
   	    Y^{T}_{N} \frac{1}{M_{RR}} Y^{}_{N} v_{2L} v_{2R} & Y^{}_{N} \frac{1}{M_{RR}} Y^{ T}_{N} v^2_{2R}
\end{array} \right) \,. \nonumber \\
 &=&\left(\begin{array}{cc}
		M_L  & M_D \\
   	    M^T_D & M_R
\end{array} \right) 
\label{eqn:numatrix}       
\end{eqnarray}
In the limit $M_L \ll M_D \ll M_R$, the type-I seesaw contribution to 
light neutrino mass is given by
\begin{eqnarray}
M^{\rm I}_\nu = - M^T_D \frac{1}{M_{R}} M_D\,, 
\end{eqnarray}
and the light-heavy neutrino mixing is proportional to $M_D/M_R \approx v_{2L}/v_{2R}$. 
With $v_{2L} \approx 174~$GeV and $v_{2R}$ around few TeV, we find that light-heavy neutrino 
mixing is large of the order of $\leq 0.1$. This large value of light-heavy neutrino mixing where 
heavy neutrinos are fixed at few keV scale, can contribute to $0\nu\beta\beta$ significantly. 

\begin{figure}[t!]
	\includegraphics[width=0.51\textwidth]{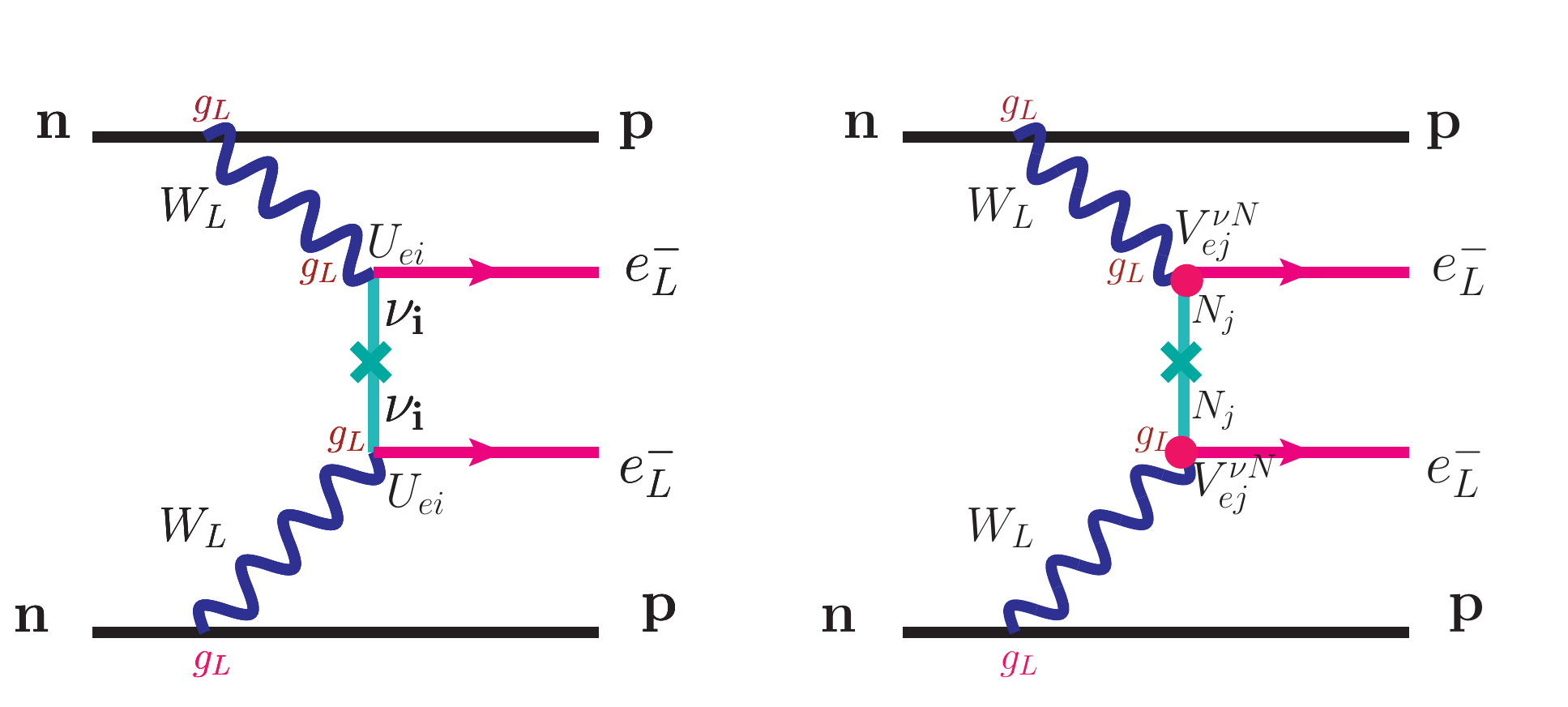}
	\caption{$0\nu\beta\beta$ decay diagrams due to purely left-handed charge current 
	interaction and with the exchange of $\nu_L$ and $N_R$.} 
	\label{feyn4}
\end{figure}
\subsection{Purely left-handed current effects}
The new physics contributions to $0\nu\beta\beta$ decay arising from purely left-handed currents 
due to exchange of keV scale right-handed neutrinos results in the following effective mass 
parameter,
\begin{eqnarray}
&&{\large \bf  m}_{\rm ee,LL}^{N} = \sum_{i=1}^3 {\mbox{V}^{\nu N}_{e\,i}}^2\, M_{i}      
\end{eqnarray}
here $M_i$ is in keV range and $\mbox{V}^{\nu N}$ is the light-heavy neutrino mixing. Since 
$\mbox{V}^{\nu N} \propto v_{2L}/v_{2R} \approx 0.01$, the effective mass parameter for $0\nu\beta\beta$ 
is estimated to be ${\large \bf  m}_{\rm ee,LL}^{N} = (0.01)^2 \cdot 10^{3}~$eV which is of the order of $0.1~$eV, 
saturating the KamLAND-Zen~\cite{KamLAND-Zen:2016pfg} bound.

\begin{figure}[h!]
	\includegraphics[width=0.51\textwidth]{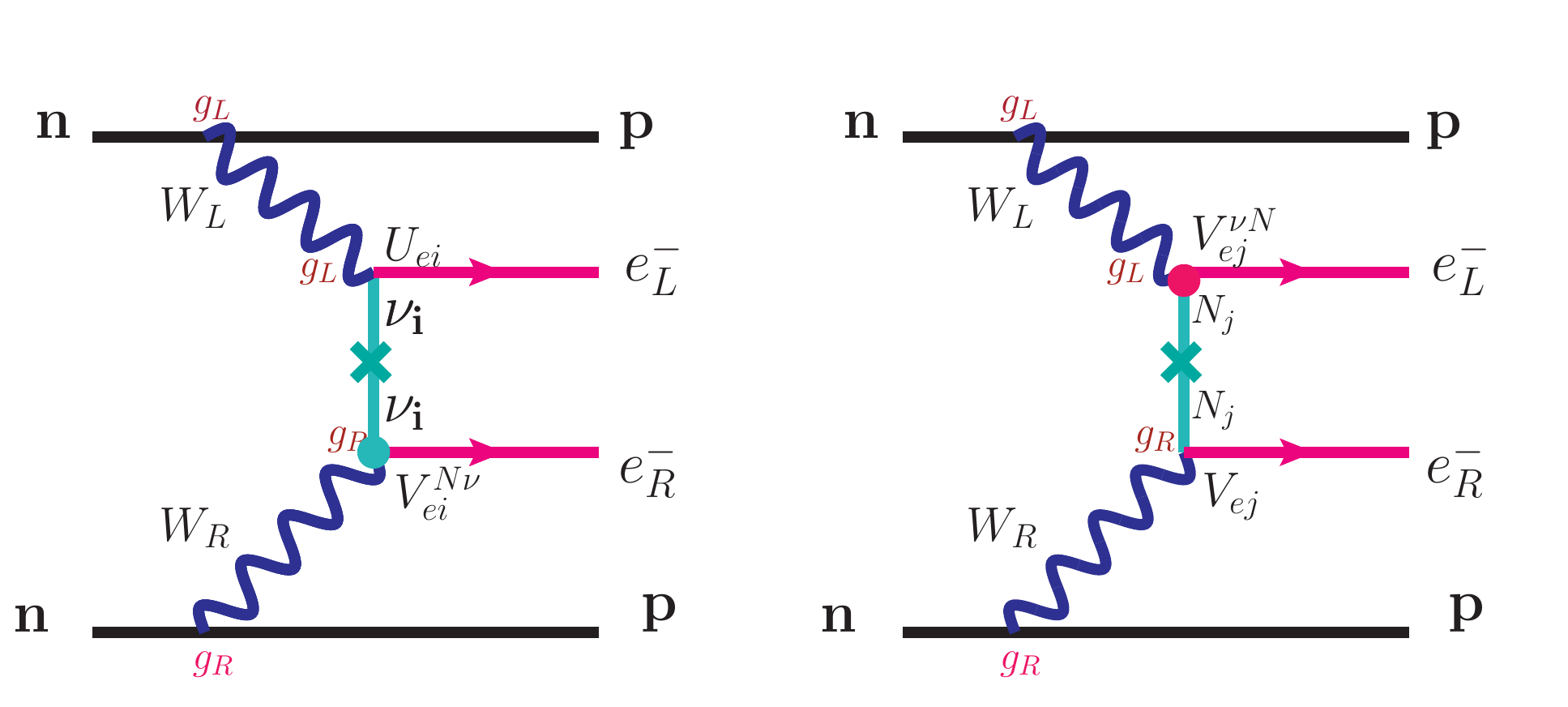}
	\caption{$0\nu\beta\beta$ decay diagrams due to mediation of one $W_L$ and 
	one $W_R$ which also involves light-heavy neutrino mixing.} 
	\label{feyn4}
\end{figure}
\subsection{From $\lambda-~$diagram}
The effective mass parameters due to the $W_L-W_R$ mediated diagrams (known as $\lambda$ diagrams) shown in figure \ref{feyn4} are given by
\begin{eqnarray}
&&{\large \bf  m}_{\rm ee,\lambda}^{\nu} = 10^{-2}\, \left(\frac{M_{W_L}}{M_{W_R}} \right)^2 
            \sum_{i=1}^3  U_{e\,i} \mbox{V}^{N \nu}_{e\,i}\, |p| \nonumber \\
&&{\large \bf  m}_{\rm ee,\lambda}^{N}=10^{-2}\, \left(\frac{M_{W_L}}{M_{W_R}} \right)^2 
            \sum_{j=1}^3  \mbox{V}_{e\,j}\, \mbox{V}^{\nu N}_{e\,j} \, |p|          
\end{eqnarray}
With $M_{W_L} \approx 80.4~$GeV, $M_{W_R} \approx 4~$TeV and $\mbox{V}^{\nu N} \simeq 0.01$, the 
effective mass parameters due to these $\lambda$ diagrams are found to be around sub-eV which can translated 
to a life-time of $10^{26}$ yrs. This value is very close to experimental bound for $\text{Xe}$ isotope. It is interesting that, such large observable heavy-light neutrino mixing arises naturally in the model without any fine-tuning of the Yukawa couplings involved. Although we consider one contribution at a time in the above discussions, in general one has to include all the contributions and at the same time keeping the light neutrino mass and mixing in the allowed range. We intend to perform a detailed study, considering the most general neutrino mass formula $M_{\nu} = M_L + M^{\rm I}_\nu$ to an upcoming work.

\section{Symmetry Breaking Pattern}
\label{sec:symbreak}
\begin{figure}
\hspace*{-0.3cm}
\includegraphics[width=1.1\linewidth]{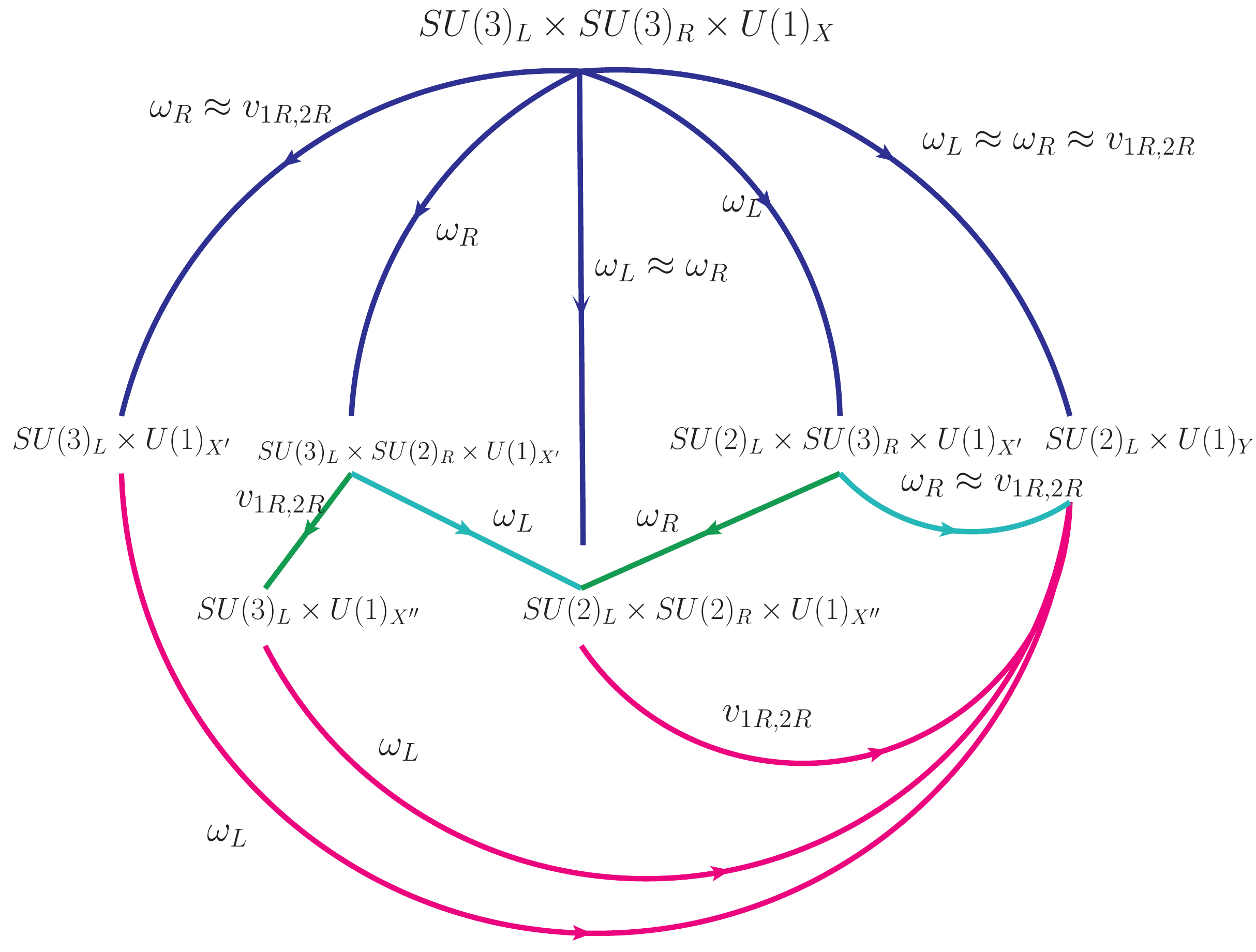}
\caption{Symmetry breaking patterns of $SU(3)_c \times SU(3)_L \times SU(3)_R \times U(1)_X$ 
        gauge symmetry to the Standard Model gauge symmetry $SU(3)_C \times SU(2)_L \times U(1)_Y$.}
\label{fig2}
\end{figure}

Depending on the scale of different vev's mentioned above, the gauge symmetry of the model can be broken 
down to that of the Standard Model through several possible symmetry breaking chains. They are summarized 
pictorially in figure \ref{fig2}. The relevant scalar potential for the model can be written as 
\begin{eqnarray}
&&V(\phi, \chi)= V_{\chi} + V_{\phi} + V_{\chi \phi}  \nonumber \\
V_{\chi}&=&\mu^2_{\chi} \left( (\chi^{\dagger}_L \chi_L) + (\chi^{\dagger}_R \chi_R) \right)+\lambda_{\chi} 
\left( (\chi^{\dagger}_L \chi_L)^2 + (\chi^{\dagger}_R \chi_R)^2 \right)\nonumber \\ &+& \lambda'_{\chi} (\chi^{\dagger}_L 
\chi_L) (\chi^{\dagger}_R \chi_R) \nonumber \\
V_{\phi}&=& \mu^2_{\phi} \left( (\phi^{\dagger}_L \phi_L) + (\phi^{\dagger}_R \phi_R) \right)+\lambda_{\phi} 
\left( (\phi^{\dagger}_L \phi_L)^2 + (\phi^{\dagger}_R \phi_R)^2 \right) 
\nonumber \\ &+& \lambda'_{\phi} (\phi^{\dagger}_L 
\phi_L) (\phi^{\dagger}_R \phi_R)\nonumber \\ 
V_{\chi \phi}&=&\rho_{\phi \chi} \left( (\chi^{\dagger}_L \chi_L) + (\chi^{\dagger}_R \chi_R) \right) 
\left( (\phi^{\dagger}_L \phi_L) + (\phi^{\dagger}_R \phi_R) \right) 
\nonumber \\ &+& \rho'_{\phi \chi} \left ( \epsilon_{ijk} 
\phi^i_L \phi^j_L \chi^k_L +\epsilon_{ijk} \phi^i_R \phi^j_R \chi^k_R+\text{h.c.} \right)  \nonumber 
\end{eqnarray}

In the scalar potential written above, the discrete left-right symmetry is assumed which ensures the equality 
of left and right sector couplings. However, as shown in earlier works \cite{univSeesawLR} in the context 
of usual LRSM with universal seesaw that the scalar potential of such a model with exact discrete left-right 
symmetry is too restrictive and gives to either parity preserving $(v_L =v_R)$ solution or a solution with 
$(v_R \neq 0, v_L = 0)$ at tree level. While the first one is not phenomenologically acceptable the latter 
solution can be acceptable if a non-zero vev $v_L \neq 0$ can be generated through radiative corrections 
\cite{ALRM}. While it may naturally explain the smallness of $v_L$ compared to $v_R$, it will constrain 
the parameter space significantly \cite{ALRM}. Another way of achieving a parity breaking vacuum is to consider 
softly broken discrete left-right symmetry by considering different mass terms for the left and right sector 
scalars \cite{lrsmpot, univSeesawLR}. As it was pointed out by the authors of \cite{lrsmpot}, such a model which 
respects the discrete left-right symmetry everywhere except in the scalar mass terms, preserve the \textit{naturalness} 
of the left-right symmetry in spite of radiative corrections. Another interesting way is to achieve parity breaking 
vacuum is to decouple the scale of parity breaking and gauge symmetry breaking by introducing a parity odd singlet 
scalar \cite{Chang:1983fu}. While we do not perform a detailed analysis of different possible symmetry breaking 
chains and their constraints on the parameter space of the model, we outline them pictorially in the cartoon 
shown in figure \ref{fig2}. As can be seen from figure \ref{fig2}, there are seven different symmetry breaking 
chains through which the gauge symmetry of the model $SU(3)_c \times SU(3)_L \times SU(3)_R \times U(1)_{X}$ 
can be broken down to that of the SM as summarized below.
\begin{itemize}
\item One step breaking: The vev's satisfy $v_{1L,2L} \ll v_{1R,2R} \approx \omega_{L} \approx \omega_R$ in this case. 
\item Two step breaking: The vev's satisfy either $v_{1L,2L} \ll \omega_L \ll v_{1R,2R} \approx \omega_{R}$ 
or $v_{1L,2L} \ll  v_{1R,2R} \ll \omega \approx \omega_{R}$ or $v_{1L,2L} \ll \omega_R \approx v_{1R,2R} 
\ll \omega_L$. The usual $331$ model presumes an intermediate stage in the first case while the usual LRSM 
or $3221$ symmetry arises an intermediate symmetry in the second case. In the third case, the $3231$ symmetry 
assumes an intermediate stage. The phenomenology of such asymmetric LRSM was discussed recently by \cite{3331750}.
\item Three step breaking: This is possible in three different ways when the vev's satisfy $v_{1L,2L} 
\ll v_{1R,2R} \ll \omega_R \ll \omega_L$ or $v_{1L,2L} \ll v_{1R,2R} \ll \omega_R \ll \omega_L$ or 
$v_{1L,2L} \ll \omega_L \ll v_{1R,2R} \ll \omega_R$. One can have both the usual LRSM or $3221$ or 
asymmetric LRSM ($3321$ or $3231$) or the usual $331$ model as an intermediate stage.
\end{itemize}

All these different symmetry breaking chains can not only provide a different phase transition history 
in cosmology but also give rise to different particle spectra including gauge bosons as well as neutral 
fermions which could be tested in different experiments.


\section{Conclusion}
\label{sec:conclude}
We have demonstrated a class of left-right symmetric model with extended gauge group 
$SU(3)_c \times SU(3)_L \times SU(3)_R \times U(1)_X$ with a universal seesaw mechanism 
for fermion masses and mixing and the implications for neutrinoless double beta ($0\nu\beta\beta$) 
decay. The novel feature of the model is that masses and mixing for left-handed and right-handed 
neutrinos are exactly determined by oscillation parameters and lightest neutrino mass. This forces the heavy neutrino masses to lie in keV regime if the $W_R$ mass is fixed at a few TeV. We show that for such a case, the heavy neutrino contribution to $0\nu\beta \beta$ remains suppressed compared to the usual light neutrino contribution. We also show that for such a TeV scale model, the heavy-light neutrino mixing can be quite large and can contribute substantially to $0\nu \beta \beta$ diagrams, keeping it within experimental reach. In the end we have discussed the scalar potential and possible symmetry breaking patterns that can be allowed for spontaneous breaking of the 3331 gauge symmetry to that of the standard model.

\begin{acknowledgments}
The authors would like to thank the organizers of the \textit{Indo-US Bilateral Workshop on Understanding 
the Origin of the Invisible Sector: From Neutrinos to Dark Matter and Dark Energy} during November 16-18, 2016 
at the School of Physics, University of Hyderabad, India where this work was initiated.
\end{acknowledgments}

\end{document}